\documentclass[prl,twocolumn,showpacs,preprintnumbers,amsmath,amssymb,superscriptaddress]{revtex4-1}

\usepackage{graphicx}% Include figure files
\usepackage{epsfig}
\usepackage{amsmath}
\usepackage{amssymb}
\usepackage{textcomp}
\usepackage{dcolumn}% Align table columns on decimal point
\usepackage{bm}% bold math
\usepackage{braket}
\PassOptionsToPackage{numbers,sort&compress}{natbib}
\usepackage{pdfpages}%for attaching the supplemental

\makeatletter
\g@addto@macro\normalsize{%
  \setlength\abovedisplayskip{4pt}
  \setlength\belowdisplayskip{4pt}
  \setlength\abovedisplayshortskip{4pt}
  \setlength\belowdisplayshortskip{4pt}
}
\makeatother

\newcommand{\BibitemShut}[1]{} %handles an error due to an outdated natbib

\begin{document}

\newcommand*{\WIPM}{State Key Laboratory of Magnetic Resonance and Atomic and Molecular Physics, Wuhan Institute of Physics and Mathematics, Chinese Academy of Sciences, Wuhan 430071, China}
\affiliation{\WIPM}
\newcommand*{\AFS}{Key Laboratory of Atomic Frequency Standards, Wuhan Institute of Physics and Mathematics, Chinese Academy of Sciences, Wuhan 430071, China}
\newcommand*{\UCAS}{University of Chinese Academy of Sciences, Beijing 100049, China}
\newcommand*{\CCAP}{Center for Cold Atom Physics, Chinese Academy of Sciences, Wuhan 430071, China}
%\affiliation{\MAINZ}
%\affiliation{\IQOQI}
%\affiliation{\IBK}

\title{Precision Measurement of the Quadrupole Transition Matrix Element in a Single Trapped $^{40}$Ca$^{+}$}
\author{H. Shao}\affiliation{\WIPM}\affiliation{\AFS}\affiliation{\UCAS}
\author{Y. Huang}\affiliation{\WIPM}\affiliation{\AFS}
\author{H. Guan}\email{guanhua@wipm.ac.cn}\affiliation{\WIPM}\affiliation{\AFS}
\author{C. Li}\affiliation{\WIPM}\affiliation{\AFS}
\author{T. Shi}\affiliation{\WIPM}\affiliation{\AFS}\affiliation{\CCAP}
\author{K. Gao}\email{klgao@wipm.ac.cn}\affiliation{\WIPM}\affiliation{\AFS}\affiliation{\CCAP}

\date{\today}% It is always \today, today,

\begin{abstract}
We report the first experimental determination of the $4s \ ^{2}S_{1/2} $ $\leftrightarrow $ $3d \ ^{2}D_{5/2}$ quadrupole transition matrix element in $^{40}$Ca$^+$ by measuring the branching ratio of the $3d \ ^{2}D_{5/2} $ state decaying into the ground state $4s \ ^{2}S_{1/2} $ and the lifetime of the $3d \ ^{2}D_{5/2} $ state, using a technique of highly synchronized measurement sequence for laser control and highly efficient quantum state detection for quantum jumps. The measured branching ratio and improved lifetime are, respectively, 0.9992(80) and 1.1652(46) s, which yield the value of the quadrupole transition matrix element (in absolute value) 9.737(43)~$ea_{0}^{2}$ with the uncertainty at the level of 0.44\%. The measured quadrupole transition matrix element is in good agreement with the most precise many-body atomic structure calculations. Our method can be universally applied to measurements of transition matrix elements in single ions and atoms of similar structure.
\end{abstract}
\maketitle

Due to its environmental isolation and long interrogation time, a single trapped $^{40}$Ca$^+$ ion has been used as an ideal system for developing optical frequency standards \cite{[1]} and for studying quantum information processes \cite{[2],[3],[4]}. A trapped $^{40}$Ca$^+$ ion has also been used for precision measurements to test atomic many-body theories \cite{[5],[6],[7]}. Such measurements include lifetimes of some low-lying states in $^{40}$Ca$^+$ \cite{[5],[6]}, branching ratios \cite{[8]}, and dipole transition matrix elements \cite{[9]}. There are similar developments for $^{88}$Sr$^+$ \cite{[10]}. Among many atomic properties, quadrupole transition matrix elements are of particular importance to design optical frequency standards; this is because in many ion clocks, such as $^{40}$Ca$^+$, $^{88}$Sr$^+$, and $^{138}$Ba$^+$, the clock reference lines are dipole-forbidden quadrupole transitions. Precise knowledge of quadrupole transition matrix elements is essential for characterizing relevant spectral lines. Since the existence of magic wavelengths for the $^{40}$Ca$^+$ clock transition has been demonstrated both theoretically \cite{[11],[12]} and experimentally \cite{[13]}, all-optical trapped ion clocks are feasible to be realized in the foreseeable future. One important issue for building such kind of trap is to overcome large ac Stark shifts due to use of high optical power; thus it is necessary to take higher-order effects into account by knowing the corresponding high-order transition matrix elements \cite{[14]}. In quantum information research, because of long coherence time of the ground and metastable states, selected quadrupole transitions for encoding a quantum bit of information are used to realize quantum logic techniques. In plasma physics, quadrupole transitions open an observational window into hot plasmas of low electron density \cite{[15]}. In astrophysics, quadrupole transition lines provide information on structure and physical characteristics of interstellar clouds \cite{[16],[17]}. Moreover, electric quadrupole transitions can be used to study atomic parity-violation in heavy ions \cite{[18],[19]}. Finally, comparison between experimentally measured quadrupole transition matrix elements and their corresponding theoretical values can test sophisticated relativistic atomic many-body theories.

In principle, dipole transition matrix elements in an atom or ion can be determined by measuring its ac Stark shifts when exposed to a light, using the concept of magic-zero (tune-out) wavelengths, such as the work on $^{87}$Rb \cite{[20]} and on $^{40}$Ca$^+$ \cite{[13]}. Dipole transition matrix elements can also be determined by comparing measurements of dispersive and absorptive light ion interactions \cite{[9]}. However, for quadrupole transition matrix elements, there have been no experimental measurements for any alkali-metal-like ions, to our knowledge. In this Letter, we report, for the first time, a determination of the quadrupole transition matrix moment (in absolute value) in $^{40}$Ca$^+$ involving the metastable $3d \ ^{2}D_{5/2} $ state and the ground $4s \ ^{2}S_{1/2} $ state by measuring the $3d \ ^{2}D_{5/2} $ state lifetime and the branching ratio of $3d \ ^{2}D_{5/2} $ decaying into $4s \ ^{2}S_{1/2} $, based on the relationship \cite{[21]} among the branching ratio $\Gamma_{ki}$, the lifetime $\tau_{k}$ , and the transition rate $A_{ki}$, i.e., $ \Gamma_{ki}=\tau_{k} A_{ki} $, where index $k$ designates the upper $3d \ ^{2}D_{5/2} $ state and $i$ the lower $4s \ ^{2}S_{1/2} $ state. The transition rate is further related to the quadrupole transition matrix moment $S^{E2}_{ki}$ by $ A_{ki} = \frac{1.11995 \times 10^{18}}{\lambda^{5}g_{k}}\vert S_{ki}^{E2}\vert^{2}$, where $ g_{k}$= 6 and $\lambda$ is the transition wavelength in \AA \cite{[1]}.

\begin{figure}[t]\begin{center}
\includegraphics[height=0.42\textwidth, width=0.45\textwidth]{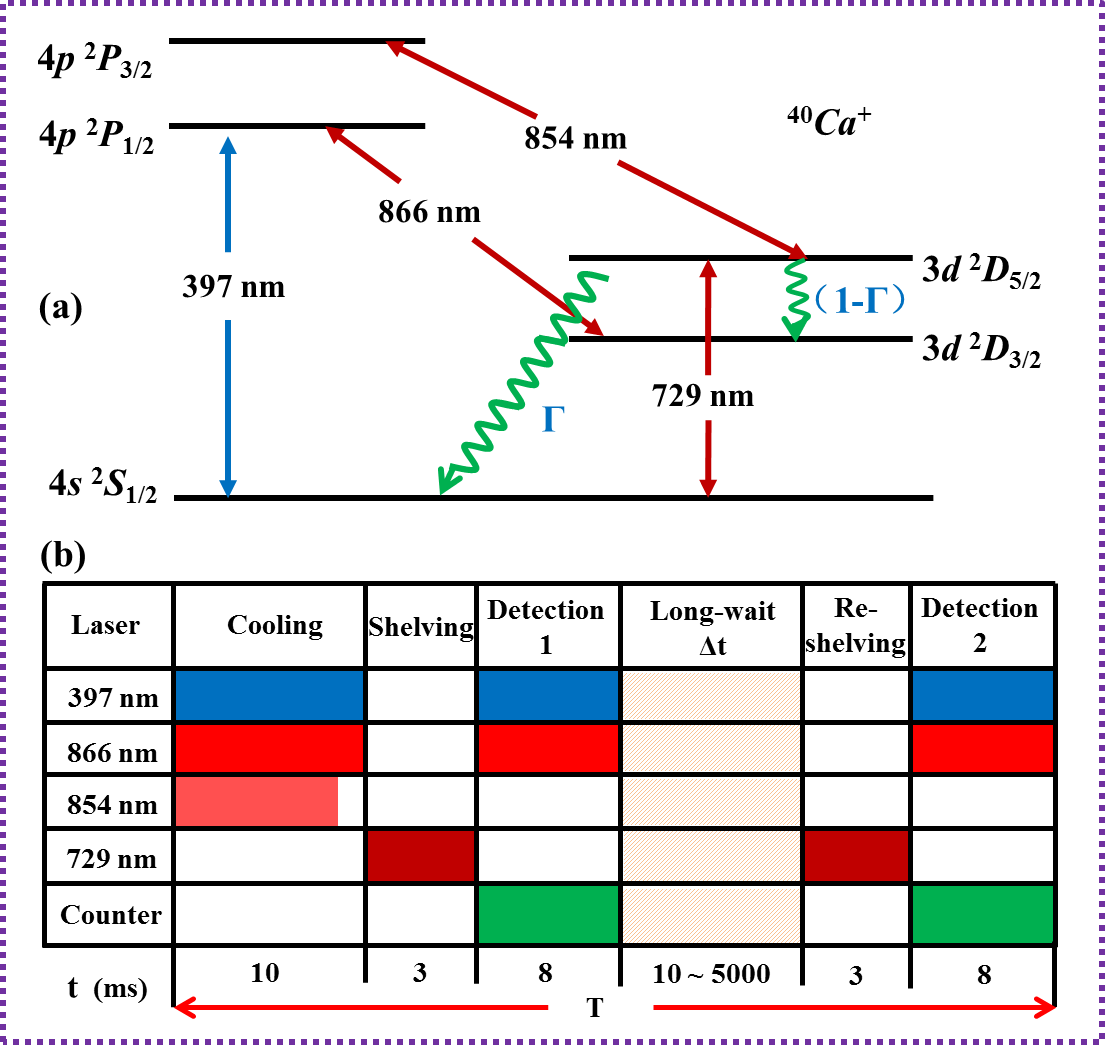}
\caption{ (Color online) (a) Partial energy-level diagram of $^{40}$Ca$^+$. The 397-nm and 866-nm lasers are for cooling and detection, the 854-nm laser for quenching, and the 729-nm laser for shelving. (b) Simplified measurement sequence.}\label{Fig1}
\end{center}
\end{figure}

Single $^{40}$Ca$^+$ ion is first laser cooled and trapped in a miniature ring Paul trap. A high-efficiency quantum state detection method for fluorescence counting and a highly synchronized measurement sequence for laser control are adopted. Figure 1 (a) shows the relevant energy levels and transition lines. Once the ion is shelved to the $3d \ ^{2}D_{5/2}$ state, it can spontaneously decay into lower-lying states with a long-wait time $\Delta t$. The decay probability $P$ obeys the exponential law $P = e^{-\Delta t/\tau_{5/2}}$, where $ \tau_{5/2} $ is the $3d \ ^{2}D_{5/2}$ state lifetime. Meanwhile, once the ion is shelved to $3d \ ^{2}D_{5/2}$, it can decay not only to the $4s \ ^{2}S_{1/2}$ ground state with the branching ratio $\Gamma$ but also to the $3d \ ^{2}D_{3/2} $ state with the branching ratio 1-$\Gamma$ \cite{[22]}. Figure 1 (b) shows a simplified sequence for the lifetime and branching ratio measurements. The branching ratio sequence consists of five steps: cooling, shelving, long-wait $\Delta t$ for decay, re-shelving, and state detection. In the first step, both the 397-nm and 866-nm lasers are executed for 10 ms to cool the ion to the Lamb-Dicke regime, and the 854-nm laser is then applied for first 8 ms in each circle to quench the ion still staying in $3d \ ^{2}D_{5/2}$ once the whole sequence is ended, which can ensure that the ion is almost in the $4s \ ^{2}S_{1/2}$ state. In the second step, the 729-nm laser is applied for 3 ms to trigger a quadrupole transition $4s \ ^{2}S_{1/2} $ $\rightarrow$ $3d \ ^{2}D_{5/2}$ so that the ion can be shelved to the $3d \ ^{2}D_{5/2}$ state, followed by an 8-ms pulse for state detection. Once the ion is coherently shelved to $3d \ ^{2}D_{5/2}$, the fluorescence at 397-nm can decrease to the background level, which will be regarded as a valid measurement; otherwise the next circle starts. In the third step, when the shelving is successfully detected, a long-wait period $\Delta t$ is performed, during which the ion decays spontaneously with no lights disturbing the process. In the fourth step, after a long-wait time $\Delta t$ for decay, the 729-nm laser is applied for 3 ms so that the ion decaying into $4s \ ^{2}S_{1/2}$ can be re-shelved. Finally, the quantum state detection is performed and the process is to observe the fluorescence signal of the ions by turning the 397 nm and 866 nm lasers on. We need to note that, all time intervals and lasers' power described above are the best parameters selected by many corresponding measurements independently. For those ions staying in $3d \ ^{2}D_{5/2}$, or decaying into $4s \ ^{2}S_{1/2}$ first followed by a re-shelving to $3d \ ^{2}D_{5/2}$, the fluorescence signal shows ``dark''; otherwise it appears ``bright''. The final probability $P_{dark}$ that the ion is in the $3d \ ^{2}D_{5/2}$ state at the end of the procedure depends on the probability $P_{5/2}$ of the decay from $3d \ ^{2}D_{5/2}$ during the wait time, the branching ratio $\Gamma$ from $3d \ ^{2}D_{5/2}$ to $4s \ ^{2}S_{1/2}$, the re-shelving efficiency $P_{re-sh}$ by the 729-nm laser, and the probability $P_{3/2}$ for the decay from $3d \ ^{2}D_{3/2}$ to $4s \ ^{2}S_{1/2}$ during the wait time. Finally, the probability that the ion will be found in the dark state can be determined from the following three procedures. The first one is the probability $P_{5/2}$ of the ion all staying in $3d \ ^{2}D_{5/2}$ and not decaying during a certain delay time $\Delta t$. The second one is the probability (1-$ P_{5/2} $)$ \Gamma $ $ P_{re-sh} $ of the ion decaying into $4s \ ^{2}S_{1/2}$ while re-shelved to $3d \ ^{2}D_{5/2}$ by a 729-nm laser after a certain delay time $\Delta t$. The third one is the probability (1-$ P_{5/2} $)(1-$ \Gamma $) (1-$ P_{3/2} $)$ P_{re-sh} $ of the ion decaying into $3d \ ^{2}D_{3/2}$ and then into $4s \ ^{2}S_{1/2}$ and finally is re-shelved to $3d \ ^{2}D_{5/2}$ by the 729-nm laser after the same delay time $\Delta t$. But if the ion always stays in $3d \ ^{2}D_{5/2}$ and does not decay into other states, when acted by the re-shelving 729-nm laser, it can induce a stimulated radiation to $4s \ ^{2}S_{1/2}$ with the probability $P_{sr}$. We can thus establish the following equation:
\begin{gather}
P_{dark} = P_{5/2}(1-P_{sr})+(1-P_{5/2})\Gamma P_{re-sh}+ \notag\\
          (1-P_{5/2})(1-\Gamma)(1-P_{3/2}) P_{re-sh}
\end{gather}
Solving for $\Gamma$ yields
$$\Gamma=\frac{P_{dark}-P_{5/2}(1-P_{sr})-(1-P_{5/2})(1-P_{3/2})P_{re-sh}}{(1-P_{5/2})P_{3/2}P_{re-sh}}\notag\\
          \eqno{(2)}$$
          
In order to extract the branching ratio $\Gamma$ from the experimental data we need to measure independently the probabilities $P_{5/2}$, $P_{3/2}$, $P_{dark}$, $P_{sr}$, and $P_{re-sh}$ at a certain dwell time $\Delta t$. Here, we choose $\Delta t$=100 ms. The final probability $P_{dark}$ is acquired by the ratio of two quantum jump numbers counted, respectively, in the second and first detection periods. The probability $P_{5/2}$ is the rate of the ion staying in $3d \ ^{2}D_{5/2}$ when $\Delta t$ =100 ms, obtained by the sequence but not including the re-shelving pulse. The probability $P_{sr}$ is also measured by the same sequence but without the wait period. However, in the first detection interval, the ion will spontaneously radiate which will affect the stimulated emission. Our final result is corrected for this effect. $P_{3/2}$ is the probability of the ion staying in $3d \ ^{2}D_{3/2}$ at the same dwell time $\Delta t$. The re-shelving probability $P_{re-sh}$ is the rate of coherently exciting the ion from $4s \ ^{2}S_{1/2}$ to $3d \ ^{2}D_{5/2}$ when the ion is re-acted by the 729-nm laser. Here, the sequence includes cooling, wait time period $\Delta t$, shelving, and detection. Introducing the wait time period $\Delta t$ is to keep the same experimental conditions as the ones for the $P_{dark}$ measurement. However, the re-shelving probability related to the branching ratio is easily affected by the laser's frequency shift and by magnetic field perturbation. In order to maintain a stable re-shelving probability, we need to minimize the magnetic field at the position of ion by using the Hanle effect. The final 10 Zeeman spectral lines are measured with full width of about 500 Hz. Also, a Ti: sapphire laser (Coherent Inc; MBR110) at 729 nm is locked to a high-finesse ULE cavity using the Pound-Drever-Hall scheme \cite{[23]}. The typical linewidth (FWHM) of the laser is $\sim $1 Hz, and the long-term drift is about 15 mHz/s \cite{[24]}. The cooling and detection lasers at 397 nm and 866 nm are frequency stabilized to the ultra-narrow linewidth 729-nm laser by a transfer cavity \cite{[25]}. To reduce the residual lights that affect our measurements, all lights are controlled by AOMs and mechanical shutters synchronously.

After having repeated 300,000 cycles of measurements, we arrive at the individual probabilities: $P_{5/2}$ = 0.91781(34), $P_{3/2}$ = 0.91979(45), $P_{dark}$ = 0.77105(13), $P_{sr}$ = 0.21307(35), and $P_{re-sh}$ = 0.59417(88), resulting in the $3d \ ^{2}D_{5/2}$ to $4s \ ^{2}S_{1/2}$ branching ratio $\Gamma$=0.9992(80). The measurement results are shown in Figure 2. 

\begin{figure}[t]\begin{center}
\includegraphics[height=0.27\textwidth, width=0.46\textwidth]{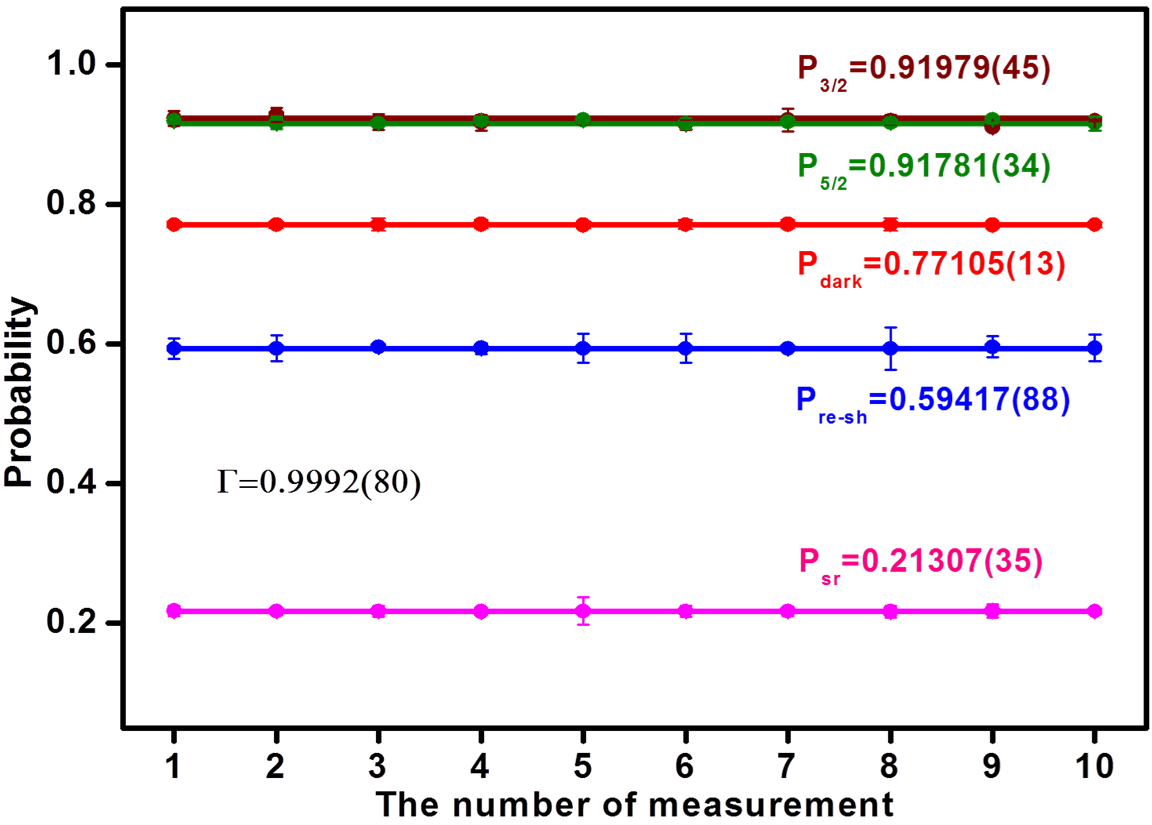}
\caption{ (Color online) Measured data points for the probabilities related to the branching ratio measurements for decay from $3d \ ^{2}D_{5/2}$ into $4s \ ^{2}S_{1/2}$. Each probability is repeated 300,000 times.}\label{Fig.2}
\end{center}
\end{figure}

For the lifetime measurement of the $3d \ ^{2}D_{5/2}$ state, the sequence consists of four steps: cooling, shelving, long wait for decay, and state detection. The procedure is similar to the measurement of the branching ratio except without the re-shelving pulse. After cooling and shelving, we need to confirm whether the ion is shelved to $3d \ ^{2}D_{5/2}$. Only in the first state detection period where the fluorescence shows a background level is regarded as a valid measurement. After a long-wait time $\Delta t$ for decay, the second state detection is performed to detect whether the ion still stays in $3d \ ^{2}D_{5/2}$ or has decayed into $4s \ ^{2}S_{1/2}$. If not, the fluorescence is kept in background level. In this experiment, for each wait time $\Delta t$, the measurement is repeated for more than 30,000 times and $\Delta t$ is set to vary from 10 ms to 5,000 ms. The spontaneous decay probability $P$ changes with $\Delta t$ and is defined as the ratio of two quantum jump numbers that are counted, respectively, in the second and the first detections. The $3d \ ^{2}D_{5/2}$ state lifetime is then determined from the exponential law $P = e^{-\Delta t/\tau_{5/2}}$. Using the measured data points shown in Figure 3 and the method of linear regression and least-squares fitting, we obtain the final lifetime $ \tau_{5/2}$=1.1653(45) s with 95\% confidence.

\begin{figure}[t]\begin{center}
\includegraphics[height=0.30\textwidth, width=0.48\textwidth]{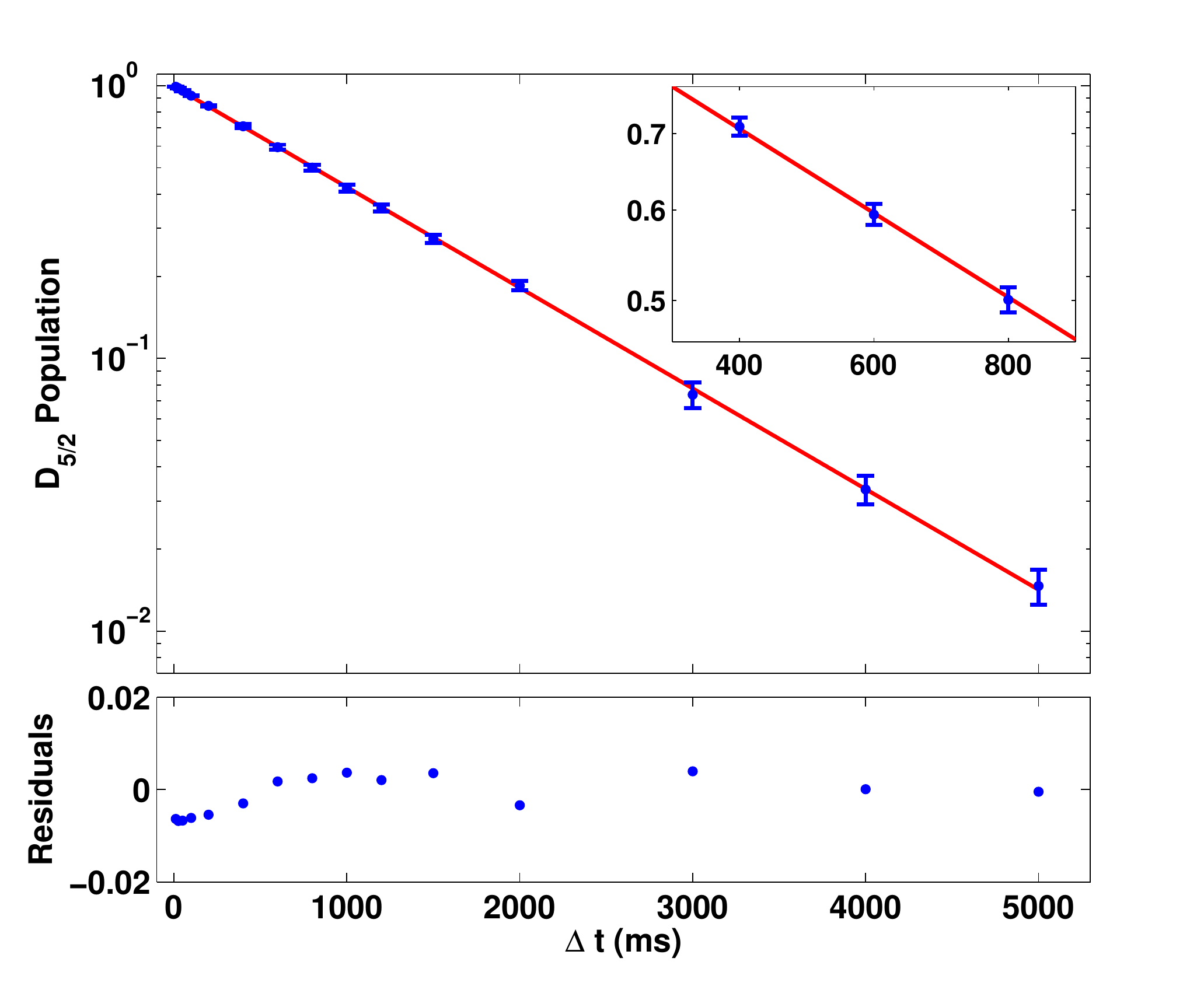}
\caption{ (Color online) Measured data points for the $3d \ ^{2}D_{5/2}$ state lifetime in $^{40}$Ca$^+$. The solid line is the fitting curve determined by the method of linear regression and least-squares to the experimental data from 10 to 5,000 ms.  Each point represents 30,000 measurements per $\Delta t$. The lower diagram shows the residuals between the data points and the fitting curve.}\label{Fig.3}
\end{center}
\end{figure}

The branching ratio and lifetime measurements are affected by many factors, such as collision, ion heating, impure composition, laser coupling, and pumping rate. The final branching ratio obtained have been corrected by considering all these factors. These physical effects can also affect the observed lifetime according to the formula \cite{[26]}: $1/\tau_{m} = 1/\tau_{nat} + \sum_{i} \gamma_{i}$, where $\tau_{m}$ and $\tau_{nat}$ are, respectively, the measured and natural lifetimes, and $\gamma_{i}$ represent contributions from other effects. In our experiment, the 397 nm and 866 nm lasers for the cooling circle are commercial semiconductor ones, which may contain other spectrum components that can induce unwanted excitations. Compared to our previous work \cite{[6]}, three improvements are made to reduce the influence on the measured lifetime. The first improvement is the use of a 406-nm narrow-band optical filter that is inputted in the 397-nm route to filter the 393-nm spectrum component contained in the 397-nm laser's spectrum. For this filter, the 393-nm laser's transmittance is less than 0.46\%, after which the residual component (the 397 nm power is about 10.2 $\mu W $, which corresponds to the intensity of 393-nm component far below 0.047 $\mu W $ after passing the filter) is unable to excite the dipole excitation transition of $4s \ ^{2}S_{1/2} $ $\to $ $4p \ ^{2}P_{3/2}$. The second improvement is the use of two 866-nm narrow-band optical filters that are inputted in the 866-nm route to filter the 854-nm component. With these two filters, the 866 nm laser's transmittance is about 81\% while the 854 nm laser's transmittance is less than 0.027\%; so the residual component (the power of 866 nm laser is about 450 $\mu W $, which corresponds to the power of 854-nm component far below 0.12 $\mu W $ after passing the filter) cannot induce the transition from $ 3d \ ^{2}D_{5/2} $ to  $4p \ ^{2}P_{3/2} $ during the detection time. The third improvement is an increase of the sample size from 5,000 to 30,000 for each delay time $\Delta t$, which greatly reduces the statistical error.

One of the main errors is the collision of the ion with the background gases in the ultra-high vacuum chamber. This process depopulates the $ 3d$  $^{2}D_{5/2}$ state      \cite{[26]}, which shortens the measured metastable lifetime. The collision rate can be monitored by recording the number of quantum jumps during the absence of the 729-nm and 854-nm lasers. With $ < 1.0\times 10^{-8}$  $ Pa $ background pressure, about 0.022(14) quantum jumps are observed per minute at different times (each procedure last 60 minutes, and repeated for 9 times), resulting in a maximum collision rate of $ 0.37(23)\times 10^{-6}$  $ms^{-1} $. The corresponding contribution to the $ 3d$  $^{2}D_{5/2}$ state lifetime is $\delta \tau _{5/2}$ = 0.5(3) ms. The heating of the ion trap also affects the measured lifetime, but through the RF-photon correlation method \cite{[27]} and selecting a suitable parameter, the excess micro-motion of the ion can be minimized. Another factor is the laser powers applied; but bringing in narrow-band optical filter in 397-nm and 866-nm routes as described above has filtered most of the unwanted components. The shelving rates also affect the measured lifetimes and branching ratio. If the 854-nm laser does not pump the ion from $ 3d$  $^{2}D_{5/2} $ to $ 4p$  $^{2}P_{3/2} $, it could decay spontaneously. When the second cycle starts, the ion will stay there for some time, which increases the measured lifetime; the corresponding shift is $\delta \tau _{5/2}$ = -1.6(6) ms.  The 397-nm shelving rate also affects the measured probability of $P_{3/2}$. We obtain the 854-nm pumping and 397-nm shelving rates 0.9981(5) and 0.9974(8) respectively under the same experimental conditions as for the above probability measurements. The wrong state detection counting, due to Poisson noise in the PMT and the spontaneous decay when detection time, which can be overcome by properly choosing a threshold value to discriminate the fluorescence of the ion in $ 4s$  $^{2}S_{1/2} $ and $ 3d$  $^{2}D_{5/2} $. The state detection error caused by the noise is $10^{-6}$ \cite{[28]}, which is ignorable; whereas the error caused by the spontaneous decay is $8.6(7)\times 10^{-4}$ that changes the measured $ 3d$  $^{2}D_{5/2} $ state lifetime by $\delta \tau _{5/2}$ = 1.0(8) ms. All factors affecting the measurements of the $ 3d$  $^{2}D_{5/2} $ lifetime are summarized in Table I. The final result for the $ 3d$  $^{2}D_{5/2} $ lifetime, after correction, is $\tau_{5/2} $ = 1.1652(46) s.

\begin{table}[htp!]
\caption{Error evaluation for the measurement of the lifetime of the $ 3d$  $^{2}D_{5/2} $ state. The detailed analysis is described in the text. The systematic errors consist of errors due to the laser's intensities, collision with background gases, heating, etc. The statistical error refers to error in data analysis.
\vspace{6pt}}
\begin{tabular}{p{4.5cm} p{2.0cm}p{1.6cm}}\hline\hline
\textbf{Effect} & \textbf{Shift(ms)}& \textbf{Unc.(ms)}  \\\hline
Fitting uncertainty & $-$ & 4.5\\
Collision depopulation & 0.5 & 0.3 \\
Heating and ion loss & $-$ & $<$0.1 \\
397-nm laser power & $-$ & $<$0.1 \\
866-nm laser power & $-$ & $<$0.1 \\
854-nm pumping rate & -1.6 & 0.6 \\
Data analysis & 1.0 & 0.8 \\\hline
Total error & -0.1 & 4.6 \\\hline\hline
\end{tabular}
\label{tbl:ErrCorr}
\end{table}

To date, the $ 3d$  $^{2}D_{5/2} $ state lifetime has been measured by many groups. In general, there exist two types of methods. One is to use laser probing technique for direct monitoring of fluorescence on the ion cloud. The other one is to use the electron shelving technique to detect single ion or ion chain quantum states. By recording the delay probability at different time delay $\Delta t$ one can obtain the lifetime. Figure 4 shows a comparison of recently measured and calculated lifetimes, where our new result has improved the best previous measured result of Ref. \cite{[6]} by a factor of 2.

With our measured branching ratio and lifetime, we can now extract, for the first time, the experimental values of the $4s \ ^{2}S_{1/2} $ $\leftrightarrow $ $3d \ ^{2}D_{5/2}$ quadrupole transition rate and transition matrix element (in absolute value), which are, respectively, 0.8575(76) $s^{-1}$ and 9.737(43) $ea_{0}^{2}$ at the levels of 0.88\% and 0.44\%. Our result for the quadrupole transition matrix element is in good agreement with the theoretical values of 9.740(47) $ea_{0}^{2}$ by Safronova \cite{[5]} and 9.713(10) $ea_{0}^{2}$ by Sahoo and Li \cite{[6]}.

\begin{figure}[t]\begin{center}
\includegraphics[height=0.3\textwidth, width=0.46\textwidth]{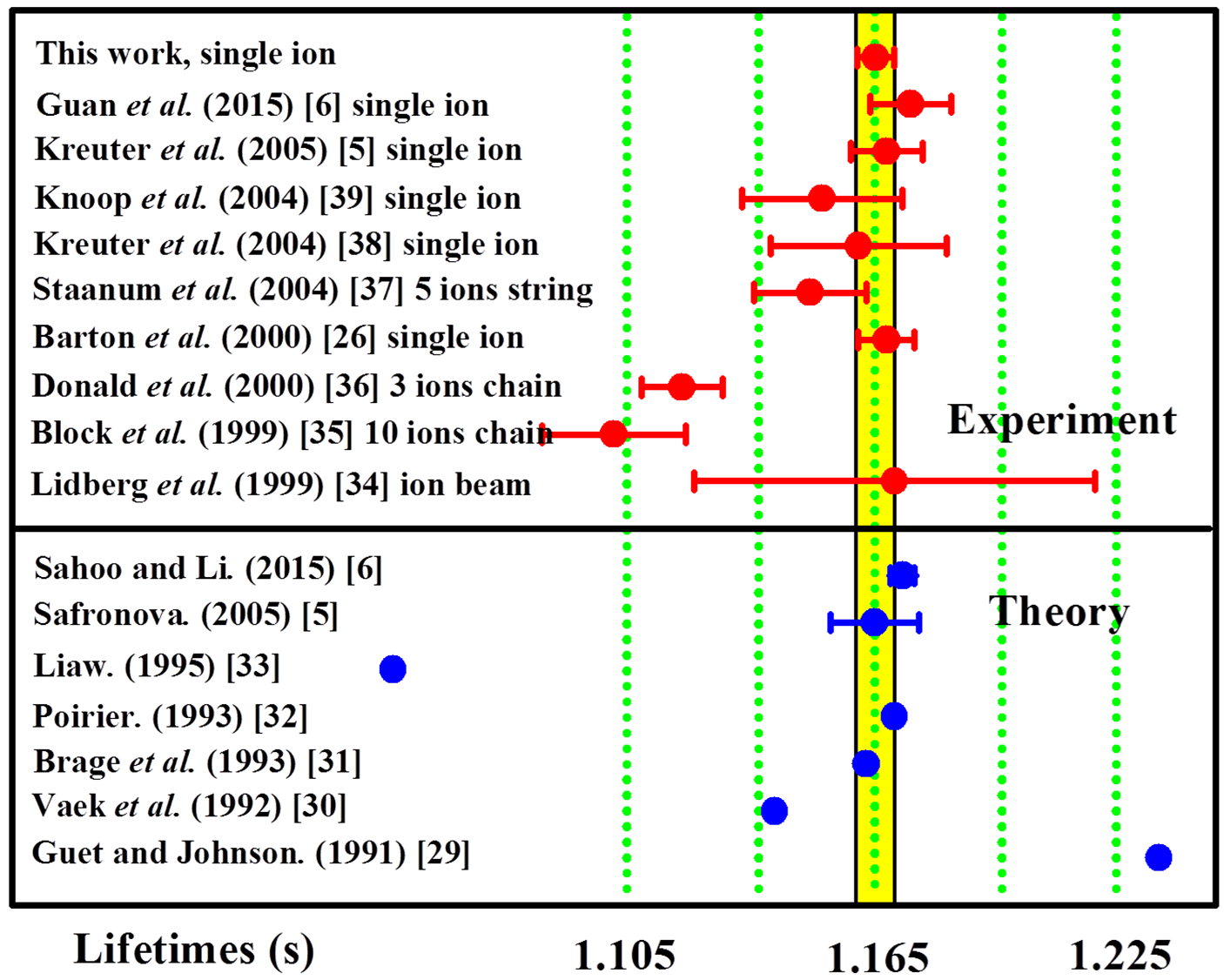}
\caption{(Color online) Comparison of the measured $ 3d$  $^{2}D_{5/2} $ state lifetime with other theoretical and experimental values.}\label{Fig.4}
\end{center}
\end{figure}

In summary, with the development of single ion manipulation technique and laser spectroscopy methods, we have measured the $ 3d$  $^{2}D_{5/2} $ state lifetime in $^{40}$Ca$^+$, which is $\tau_{5/2} $ = 1.1652(46) s, with the uncertainty of 0.4\%, a factor of 2 improvement over the best previous measurement \cite{[6]}. Meanwhile, for the first time, we have measured the branching ratio for the decay $ 3d \ ^{2}D_{5/2} $ $\to$  $4s \ ^{2}S_{1/2} $ as 0.9992(80). With these two measured results, we have extracted the first experimental value of the $4s \ ^{2}S_{1/2} $ $\leftrightarrow$ $ 3d \ ^{2}D_{5/2} $ quadrupole transition matrix element 9.737(43) $ea_{0}^{2}$, with the uncertainty at the level of 0.44\%. Our results can serve as a test bed for atomic structure calculations and for high precision laser spectroscopy \cite{[40]}. Nevertheless, the main sources of uncertainties originate from the statistical errors in the lifetime measurement, collision with background gases, and the existence of multiple decay channels in the branching ratio measurement. Further improvement in accuracy is still possible by collecting larger number of quantum jumps under a better vacuum condition.

\begin{acknowledgments}
We thank Z. Yan, A. Derevianko, and B. K. Sahoo for fruitful discussion. This work is financial supported by the National Basic Research Program of China (Grant No. 2012CB821301), by the National Natural Science Foundation of China (Grants No. 91336211, No. 11034009, No. 11474318, and No. 11304363), and by Chinese Academy of Sciences.
\end{acknowledgments}

\bibliographystyle{apsrev4-1}
\bibliography{scatter_bib}

%the following is a hack to attach the supplemental material
%includepdf for the whole, doc at once doesn't work, so
%we do it like this

\end{document}